\documentclass[
  ]
  {aipproc}

\layoutstyle{8x11double}

\SetInternalRegister\hbadness{8000} 

\newcommand\doingARLO[2][]{%
  \ifx\mmref\undefined #1\else #2\fi
}

\begin{document}

\title 
      [Swift Triggering]
      {The Trigger Algorithm for the Burst Alert Telescope on Swift}


\author{Ed Fenimore}{
  address={Los Alamos National Laboratory, Los Alamos, NM USA},
  email={efenimore@lanl.gov},
}

\iftrue

\author{David Palmer}{
  address={Los Alamos National Laboratory, Los Alamos, NM USA}
}

\author{Mark Galassi}{
  address={Los Alamos National Laboratory, Los Alamos, NM USA}
}

\author{Tanya Tavenner}{
  address={Los Alamos National Laboratory, Los Alamos, NM USA}
}

\author{Scott Barthelmy}{
  address={Goddard Space Flight Center, Greenbelt, MD USA}
}
\author{Neil Gehrels}{
  address={Goddard Space Flight Center, Greenbelt, MD USA}
}
\author{Ann Parsons}{
  address={Goddard Space Flight Center, Greenbelt, MD USA}
}
\author{Jack Tueller}{
  address={Goddard Space Flight Center, Greenbelt, MD USA}
}

\fi

\begin{abstract}
The Swift Burst Alert Telescope (BAT) is a huge (5200 cm$^2$) coded
aperture imager that will detect gamma-ray bursts in real time and provide
a location that the Swift satellite will use to slew the optical and x-ray
telescopes. The huge size of BAT is a challenge for the on-board 
triggering:
a change as small as 1\% is equivalent to a 1 $\sigma$
statistical variation in 1 second. There will be three types of triggers, 
two based on rates and one based on images. The first type of trigger is for 
short
time scales (4 msec to 64 msec). These will be traditional triggers
(single
background) and we check about 25,000 combinations of
time-energy-focal plane subregions per second. The second type of trigger
will be similar to what is used on HETE: fits to multiple background 
regions to remove
trends for time scales between 64 msec and 64 seconds. About 500
triggers will be checked per second. For these rate triggers, false
triggers and variable non-GRB sources will be rejected by requiring a new
source to be present in an image. The third type of trigger works on 
longer time scales (minutes), and will be based on routine images that are
made of the field of view. 
\end{abstract}

\date{\today}

\maketitle

\section{Introduction}

\indent \par  The Burst Alert Telescope (BAT) on Swift is a large (5200 cm$^2$) 
CZT-based coded aperture imager. BAT's primary role on the Swift satellite 
is to detect when a gamma-ray burst (GRB) starts, quickly locate it, and 
direct 
the Swift satellite to  point the optical and x-ray telescopes at 
the source. The BAT triggers must not only detect the occurrence of a GRB,
but they also must select the time periods to form an image. This requires 
the triggering code to identify a range of times (the "background" period) 
when there is no apparent emission from the GRB and a range of times (the 
"foreground" period) where the GRB probably will produce the strongest 
image. 

The BAT triggering code has three types of triggers. Two of these are 
"rate" triggers based on statistically significant increases in the 
counting rate in the focal plane (or a portion  of the focal plane), and 
one is an "image" trigger based on new significant sources found in images 
of the field of view (FOV). The rate triggers are divided into the 
"short" rate triggers (with foreground periods of less than or equal to 64 
msec) and 
"long" rate triggers (with foreground periods larger than or equal to 64 
msec). The 
image triggers are intended for longer periods of time (from 64 sec to 
many minutes). 

One goal is to explore the widest possible parameter space. As such, as 
many triggers as possible will be run simultaneously until the flight 
computer is nearly saturated. Thus, special attention will be paid to the 
CPU usage.

It is not hard to design a triggering code that responds to GRBs. The real 
challenge in the triggering code is to avoid false triggers. This is a 
special problem with BAT because its huge size means that a very slight 
trend 
can appear to be a significant increase in the count rate: a 1\% change in 
the BAT count rate between the background and foreground regions appears 
to be a 1 $\sigma$ variation for a 1 sec foreground sample. Thus, the 
chief danger is false triggers due to trends in the background, 
variations of uninteresting sources in the FOV, or minor 
configuration 
changes (such as automatic gain adjust) which produce the appearance of a 
statistically significant increase.

\section{Short Rate Triggers}

 Checking many short time scales in a triggering code can require 
most 
of the CPU time. The background counting rate of BAT is not 
expected to 
change on short time scales (i.e., less than a few seconds). Thus, for the 
short time scales we will use simple traditional triggers where there is a 
single background period of fixed duration before the foreground period. 
This is the type of trigger 
that was used on all GRB experiments from Vela to BATSE. 

The short trigger looks for statistically significant increases in the 
count rate on five time scales: 4, 8, 16, 32, and 64 msec. This is done for 
nine different regions of the focal plane (four quadrants, the left half, 
right half, top half, bottom half, and the full focal plane) and for four 
energy ranges. Thus, there are 36 combinations of focal plane regions and 
energy ranges. Within each 1.024 second period there are 256 4-msec 
samples to check, 
256 8-msec samples to check (assuming the foreground periods are checked 
at all 4-msec phases), 128 16-msec samples (assuming the foreground 
periods are checked at all 8-msec phases), 64 32-msec samples, and 32 
64-msec samples.

 Overall, there are more than 26,496 short trigger samples to check every 
1.024 second. The calculational effort is optimized by having the code 
responsible for reading the photons from the focal plane search for the 
maximum number of counts at each time scale and region-energy combination. 
Every 320 msec, the short triggering code is sent 180 samples: the maximum 
counts seen in the 5 time scales and the 36 region-energy combinations. 
The triggering code only has to check the maximum sample that occurred 
within each time scale-region-energy combination, not every observed 
sample. By having the code that ingests the photons identify the maximums, 
we can effectively check 26,000 samples a second with about 560 actual 
trigger calculations per second. 

 All of the short trigger calculations for a particular set of 180 samples 
use the same background rates based on a 1.024 sec period. These 
background rates are determined by the running sums in the long trigger 
algorithm (see below). Let 
$C_{i,k}$ be the maximum counts observed on the $2^k$ msec time scale
in the 
$i^{\rm th}$ region-energy combination. Let $B_i$ be the counts observed 
in 1024 msec for the $i^{\rm th}$ region-energy combination. The short 
trigger "score" is effectively the $\sigma^2$ of the net signal relative 
to the expected statistical variation. (We use $\sigma^2$ to avoid taking 
square roots: one can more easily find the maximum of $\sigma^2$ than the 
maximum of $\sigma$ and both methods will point to the same sample.) The 
definition of the short trigger score 
is:
\begin{equation}
S = {(C_{i,k}-B_i2^{k-10})^2 \over B_i2^{k-10}+\sigma^2_{\rm min}}~.
\label{SHORTSCORE}
\end{equation}
Here, $\sigma^2_{\rm min}$ is a commandable control variable to ensure 
that there is a minimum variance when the counts are small. A trigger is 
declared if $S$ is greater than a threshold, $\sigma^2_{\rm threshold}$. 
Each 
of the triggers for the 180 combinations are controlled by three 
commandable variables: an enable/disable, $\sigma^2_{\rm threshold}$, and 
$\sigma^2_{\rm min}$.

Once a short trigger exceeds its threshold, the code will search stored
data within the 320 msec period to find when the identified exceedance 
occurred. That period of
time becomes the foreground period for the imaging and the background is 
taken from the most recent 8 sec focal plane accumulation.

\section{Long Rate Triggers}

To avoid trends on longer time scales, one must fit a function to the  
background and remove the trend. This is the technique pioneered by 
the HETE GRB trigger (\cite{fenhetetrig}[Tavenner, 
et al. 2002]). For each 
trigger 
criterion, we specify starting and ending times for up to three background 
regions and a foreground region. The background regions can
either be all before the foreground samples (an extrapolation) or
can bracket the foreground sample (an interpolation). A goal of Swift is 
to rapidly determine locations, so it would seem that having background 
regions after the foreground would only delay the location 
determination. However, most likely, a GRB will trigger one of the short 
traditional triggers or trigger a long extrapolation trigger  
with a threshold set high enough to suppress false triggers due to trends. 
The interpolation triggers only come into play to detect the GRB when the 
other types of 
triggers have already failed. The other crucial use of the extrapolation 
triggers is to identify the best foreground/background combination after 
the event has been detected. The trigger algorithm continues to 
process the time series 
after the 
initial detection in order to find the overall maximum trigger score 
and use the 
resulting foreground/background periods until a new source is 
found in the images.

The long rate triggers are based on time series with 64 msec time
resolution and are much more complicated than the short rate triggers. To
cover as wide as parameter space as possible, the triggering code must be
very efficient. In previous trigger algorithms (such as HETE), most of the
CPU usage is for forming sums of counts. For each long trigger, one needs
to have the sum of counts in several background periods (probably
involving a few hundred 64 msec samples) plus the counts in foregrounds
ranging from 64 msec in duration up to perhaps 64 sec. The trick to an
efficient triggering algorithm is to store the integral sum of counts, not
the counts within samples. The BAT triggering code maintains 36 time
series in circular buffers (for the 9 detector regions and 4 energy
ranges). Each time series consists of the integer sum of the counts from
when the instrument was turned on up to time $T$: 
\begin{equation}
I_i(T) = \sum_0^T C_i~. \label{TIMESERIES} 
\end{equation} 
To obtain the
sum of counts within a particular time period (say $T_1$ to $T_2$), one
needs only a single arithmetic step, $I_i(T_2)-I_i(T_1)$, rather than a
sum over the samples between $T_1$ and $T_2$. (Eventually, the integer sum
will overflow the register capability of the computer. This is 
accommodated by adding the maximum range of the computer registers if the 
difference
is negative.)

A second integral sum adds up the number of invalid samples for each of 
the 36 region-energy combinations. For example, if there is a 
configuration change in one of the quadrants, a 1 is added to the 
corresponding invalid integral sum. Whenever we seek the number of counts 
from a period of time that includes that sample, we also check that the 
difference 
of the integral sum of invalids that cover that same duration is zero. 
For every time that  the high voltage is off or something else disables the 
detectors, a 1 is added to the sum of invalid samples. The trigger code 
runs all the time and each criterion turns on as soon as the corresponding 
difference in the sum of invalid samples is zero. Thus, criteria that 
require fewer samples turn on as soon as they are ready, increasing the 
on-time for the triggers.

 Each long rate trigger is controlled by about 30 commandable parameters.
These parameters define the relative times of several background periods,
the time, duration, and amount to step the phase for the foreground
period, the degree of the polynomial to fit, a minimum variance
($\sigma_{\rm min}$, needed for low count rates), a systematic noise level
($\beta$, needed for high count rates), a threshold for declaring a
trigger ($\sigma^2_{\rm threshold}$), several parameters that control the
CPU usage, and an enable/disable parameter. 
The CPU
usage is controlled in several ways. One can specify which tick of the 64
msec clock that each criteria is evaluated. This ensures that the CPU
usage is evenly spread out. There is a CPU usage monitor and if a
commandable level is exceeded, triggers will autonomously turn themselves
off for a commandable period of time to maintain an acceptable CPU usage.

To obtain the long rate trigger score, one first finds the counts and 
variance on the counts in the foreground period ($C_{\rm fore}, 
\sigma^2_{\rm fore}$). Second, one fits a function 
(constant, linear, 
$2^{nd}$ order) to the background samples. Third, one integrates the fit 
function during the foreground sample to find the expected background rate 
($C_{\rm back}, \sigma^2_{\rm back}$). (We use "model variance", so 
$\sigma^2_{\rm 
fore}$ is actually based on $C_{\rm back}$.) Finally, the trigger score 
is calculated as:
\begin{equation}
S = {(C_{\rm fore}-C_{\rm back})^2 \over \sigma^2_{\rm 
fore}+\sigma^2_{\rm back}+\sigma^2_{\rm min}+ \beta^2 C^2_{\rm back}}~.
\label{LONGSCORE}
\end{equation}
Here, $\sigma^2_{\rm min}$ provides a minimum variance to protect against 
very low counts and $\beta$ protects against systematic effects at high 
count rates. When the count rate is very high, the variations will no 
longer be Poissonian. The presence of $\beta$ converts the trigger score 
from a signal-to-noise criteria, to a fractional difference criteria. The 
units on $\beta$ is fractional change per effective $\sigma$. At high 
count 
rates, the 
$C^2_{\rm back}$ term will dominate over the $\sigma^2$ terms, and the 
trigger score becomes 
\begin{equation}
S =  \bigg[ { {C_{\rm fore}-C_{\rm back} \over C_{\rm back} } \over 
\beta}\bigg]^2~.
\label{LONGSCOREBETA}
\end{equation}
For example, if the foreground has a  5\% net increase over the background 
and $\beta = 0.025$ (i.e., 2.5\%), then the trigger score will be 
$\approx 2^2$. 
If one did not have the $\beta$ term, a 5\% net increase would be about 5 
$\sigma$ (assuming the BAT background rate of $~10$ KHz) and the trigger 
score would 
be $5^2$. 

The huge size of BAT means that even a small change in a non-transient, 
but variable  source 
will appear to be a statistically significant change in the count rate. 
Sources such as Cyg X-1, Her X-1, and Sco X-1 could easily produce 
constant triggers forcing us to raise our thresholds. To guard against 
this, we will implement several new concepts. 
Both are derived from the  coded aperture technique "URA-tagging" 
(\cite{fen87}) which  effectively deconvolves the mask pattern in real 
time for selected source locations. Each photon is 
assigned a weight factor that is related to the probability that it came 
from a
designated source location, that is, whether it could reach the detector
through the mask pattern. The sum of the probabilities as a 
function of time becomes that
source's strength at that time. Background can be removed by scaling the 
weight factors to be negative for regions totally blocked. Effectively, 
the 
sum of the counts in the detectors that cannot see the location is 
subtracted from the locations which can.
Although developed for URA patterns, the
mask tagging also works for random patterns such as BAT. 

One use of the mask tagged rates is as a 
"veto" trigger.
Some of the 
trigger criteria will process 
the mask-tagged time histories in the same way that the 36 region-energy 
combinations 
are processed (i.e., they will be stored as integral sums like  equation 
\ref{TIMESERIES}). If the trigger criteria exceeds a threshold, a 
(commandable) set of regular triggers are disabled for that time period. 
For example, we might have veto triggers that have foreground durations of 
1, 5, and 10 sec that are applied to the tagged time series for Cyg X-1. 
It any of those veto triggers exceed their thresholds, we disable the 1, 
5, and 10 sec long rate criteria. We will have the capability to 
simultaneously track up 
to three mask-tagged sources for the veto system.

The second use of the mask tagged rates will be to correct trigger series 
by subtracting out the variations seen by the mask tagged sources. This 
raises the noise levels so the vetoing system is better if there are only 
a few excursions to accommodate.

We plan space for about 500 long rate trigger criteria. The parameters 
that control the CPU usage ensure that we will be able to run as many as 
possible.

\section{Image Triggers}

To search for GRBs or other transients on long time scales, we will form 
images of the FOV and search for new objects. The flight software forms an 
on-board encoded image every 8 seconds which are used as the background 
images for the long rate triggers. Those images are combined together on 
three different time scales (perhaps 64 sec, 10 minutes, half an orbit) 
and the on-board image deconvolves the mask pattern. (The on-board image 
process consists of a non-iterative clean to remove known bright 
sources, a mask deconvolution, and a back project analysis to refine the 
location.) We search each such 
image for significant sources and eliminate known sources using an 
on-board 
table.  Sources that exceed a commandable threshold are declared sources 
suitable for Swift to slew to.

\section{Summary}

 The BAT instrument on Swift will have trigger software that can explore a 
wide parameter space (4 msec to orbital time scales, 4 different energy 
ranges, subregions of the FOV).  About 16,000 commandable "knobs" can be 
used to optimize its performance. Of course to avoid massive confusion, 
only a few of those will actually be adjusted on orbit. The software will 
autonomously adjust the number of criteria to maximize the use of the 
available CPU power. 

We have several lines of defense against false triggers. The first is that
the hundreds of parameters that control the trigger calculations are all
commandable from the ground. The second is that we can remove trends by 
fitting a function to background samples. The third is that we will 
determine in real time the variations of sources such as Cyg X-1 and 
remove their effects from the triggering. The final line of  defense 
against false triggers
is that we will form an image and 
check to see if the "GRB"  is in the direction of a known variable source 
before slewing the satellite.

\end{document}